\documentclass[preprint,aps,tightenlines,floatfix,showpacs]{revtex4}
\usepackage{graphics}
\usepackage{bm}
\usepackage{amsmath}
\usepackage{amssymb}

\newcommand{\sfrac}[2]{\ensuremath{\textstyle\frac{#1}{#2}}}

\begin{document}

\title{Lehmer's interesting series}

\author{Freeman J. Dyson}
\affiliation{Institute for Advanced Study, Princeton, New Jersey, U.S.A.}
\email{dyson@ias.edu}

\author{Norman E. Frankel}
\affiliation{Department of Physics, University of Melbourne, 
Victoria, Australia}
\email{nef@unimelb.edu.au}

\author{M. Lawrence Glasser}
\affiliation{ Department of Physics,  Clarkson University,
Potsdam, New York, USA}
\email{laryg@tds.net}

\date{\today}

\begin{abstract}
The series
\begin{displaymath}
S_k(z) = \sum_{m=1}^{\infty} \frac{m^kz^m}{{2m \choose m}}
\end{displaymath}
is evaluated in a non-recursive and closed process and it can be 
analytically continued beyond its 
domain of convergence $0\le |z|<4$ for $k=0,1,2,\cdots$. From this we  provide 
a firm basis for  Lehmer's  observation that $\pi$ emerges from the limiting 
behavior of $S_k(2)$ as $k\rightarrow\infty$.
\end{abstract}

\maketitle

\section{Introduction}

 In the paper~\cite{Le85} marking the 60th anniversary of his first 
contribution to the American Mathematical Monthly, D. H. Lehmer studied two 
classes of interesting infinite series; interesting in the sense that the 
terms and sum are known explicitly. Lehmer evaluated the series of class
II, whose origin he attributes to L. Comtet~\cite{Co74}, namely
\begin{equation}
S_k(z)=\sum_{m=1}^{\infty}\frac{m^kz^m} {{2m  \choose m }}
\label{Coeq}
\end{equation}
for non-negative integer $k$. He did so by manipulating the Taylor expansion 
for the arcsine function, so arriving at a general formula for 
Eq.~(\ref{Coeq}) in terms of
recursively defined polynomials.  As we show below,  the domain of
convergence for Eq.~(\ref{Coeq}) is $|z|<4$. Among publications stimulated by
Ref.~\cite{Le85} are those of 
Borwein and Girgensohn~\cite{Bo05} and of Mathar~\cite{Ma09}. Also there 
are recent papers
\cite{Sh00,Ba04,Da04,Su04,Ba05,So06,Sp06,Ya06,Ya07,So10}
that contain other relevant references.  
 
Lehmer's class II series can also be expressed as  generalised
hypergeometric functions  and this has been exploited by 
Krupnikov and K\"olbig~\cite{Kr96} to construct a table of values of 
$\;_{p+1}F_p$ for unit argument and rational parameters.  Lehmer's
result,
\begin{equation}
S_2(2) = 11 + \sfrac{7}{2}\pi ,
\label{Leeq}
\end{equation}
illustrates the fascinating feature that $S_k(2)=R_1(k)+R_2(k)\pi$, 
where $R_j(k)$ is rational and, empirically, the ratio $R_1(k)/R_2(k)$ is
an approximation to $\pi$ which improves as $k$ increases. In fact, as we shall
illustrate below, in this way $S_k(2)$ appears to give $\pi$ to roughly 
$k$ decimal places. Our aims in this paper are to express $S_k(z)$ for $|z|<4$
in non-recursive form and to investigate the indicated approximation to $\pi$.

\section{Calculation}
    
We begin with the observation that $\frac{1}{m}\ C^{2m}_m = B(m,m+1)$,  
where $B$ is 
Euler's beta function and $C^N_M$ denotes the binomial coefficient.  Hence
\begin{equation}
S_k(z)=\int_0^1 \frac{dt}{t} \sum_{m=1}^{\infty}m^{k+1}[zt(1-t)]^n .
\label{Seq1}
\end{equation}
Next,  Euler's famous 1775 discovery~\cite{Eu75},  embodying the Eulerian 
triangular numbers, 
\begin{equation}
\sum_{m=1}^{\infty}\ m^p\ X^m = \sum_{n=1}^p \sum_{m=1}^n\ (-1)^{m+n}
\ C^n_m\ m^p\ X^n\ (1-X)^{-n-1} ,
\label{Eueq}
\end{equation}
  gives
\begin{equation}
S_k(z) = \sum_{n=1}^{k+1} \sum_{m=1}^n\ (-1)^{m+n}\ C^n_m\ m^{k+1}
\int_0^1 \frac{dt}{t} \frac{[zt(1-t)]^n}{[1-zt(1-t)]^{n+1}} .
\label{Seq2}
\end{equation}

 It can be shown that
\begin{equation}
\int_0^1 \frac{dt}{t} \frac{[zt(1-t)]^n}{[1-zt(1-t)]^{n+1}}
= \frac{\Gamma(n)}{\Gamma(n+\sfrac{1}{2})}\
X^n\;_2F_1(-\sfrac{1}{2},n;n+\sfrac{1}{2};-X) ,
\label{inteq}
\end{equation}
 where $X = z/(4-z)$ so
\begin{equation}
S_k(z)=
\sqrt{\pi}\ \sum_{n=1}^{k+1}\sum_{m=1}^n(-1)^{m+n}
\frac{\Gamma(n)\ C^n_m}{\Gamma(n+\sfrac{1}{2})}\ X^nm^{k+1}
\;_2F_1(-\sfrac{1}{2},n;n+\sfrac{1}{2};-X) .
\label{Seq3}
\end{equation}
This gives, for example, $S_1(2)=\pi+3$. \\

Now using Eq. [7.3.6(1)] in Ref.~\cite{Pr86}, i.e.
\begin{equation}
{\phantom{\sum_{n=1}^{k+1}}}
\;_2F_1(-{\sfrac{1}{2}}, n;n+{\sfrac{1}{2}};-1) =
2^{-n}\ \sqrt{\pi}\ \Gamma(n + \sfrac{1}{2})
\left\{ 
\left[ \Gamma(\frac{n+1}{2})\right]^{-2}
+ \frac{n}{2}\left[\Gamma(\frac{n}{2} + 1)\right]^{-2}
\right\} ,
\label{f-eqn}
\end{equation}
\begin{equation}
S_k(2) = \pi \sum_{n=1}^{k+1} \frac{n!}{2^n} E(k,n)
\left\{ \frac{(n-1)!}{\left[\Gamma\left(\frac{n+1}{2}\right)\right]^2}
+\frac{n!}{2\left[\Gamma\left(\frac{n+2}{2}\right)\right]^2}\right\} ,
\label{Seq4}
\end{equation}
where
\begin{equation}
E(k,n) = \frac{(-1)^n}{n!} \sum_{m=1}^n (-1)^m\ C^{n}_m\ m^{k+1} ,
\label{Stirs}
\end{equation}
are Stirling numbers of the second kind.
The right hand side of Eq.~(\ref{Seq4}) is manifestly of the form 
$R_1(k) + R_2(k) \pi$ and leads to Lehmer's  insightful observation, 
which we express more transparently in the form 
\begin{equation}
\lim_{k\to\infty} \frac{t_2(k)}{t_1(k)} = 2 , 
\label{Leobs}
\end{equation}
where 
\begin{align}
t_1(k) =  t_{11}(k) + t_{12}(k)
\hspace*{0.5cm}&{\rm and}\hspace*{0.5cm}
t_2(k) =  t_{21}(k) + t_{22}(k)
\\
t_{11}(k) &= \sum_{n=1}^{\left[\frac{k+1}{2}\right]}
\frac{E(k,2n)}{4^n}\ \frac{(2n)!\ (2n-1)!}
{\left[\Gamma\left(n+{\sfrac{1}{2}}\right)\right]^2}
\\
t_{12}(k) &= \sum_{n=0}^{\left[\frac{k}{2}\right]}
\frac{E(k,2n+1)}{4^{n+1}}\frac{\left[(2n+1)!\right]^2}
{\left[\Gamma\left(n+{\sfrac{3}{2}}\right)\right]^2}
\label{t12eq}
\\
t_{21}(k) &= \sum_{n=1}^{\left[\frac{k+1}{2}\right]}
\frac{E(k,2n)}{4^n}\ (2n)!\ C^{2n}_n
\\
t_{22}(k) &= \sum_{n=0}^{\left[\frac{k}{2}\right]}
\frac{E(k,2n+1)}{4^n}\ (2n+1)!\ C^{2n}_n .
\label{t22eq}
\end{align}

For large summand $n$ it is clear that the sums $t_{ij}$ are dominated by 
their "tails", but for large $n$
\begin{displaymath}
 \frac{(2n-1)!}{\left[\Gamma(n+\sfrac{1}{2})\right]^2} \
\rightarrow\ \frac{1}{2} C^{2n}_n \hspace*{0.4cm};\hspace*{0.4cm}
 \frac{1}{2}\left[\frac{(2n+1)!}{2\Gamma(n+\sfrac{3}{2})}\right]^2
\rightarrow\  C^{2n}_n .
\nonumber
\end{displaymath}
Hence, as $k\rightarrow\infty$, $2t_{11}(k)\rightarrow t_{21}(k)$ and 
$2t_{12}(k)\rightarrow t_{22}(k)$. Thereby Lehmer's assertion is demonstrated.

For $z=-2$, we have
\begin{align}
S_k(-2) &=
\sum_{n=1}^{k+1} (-1)^n\ n!\ \frac{E(k,n)}{3^n}
\nonumber\\
&\hspace*{0.3cm}\times
\left[\frac{1}{n} + \sum_{p=0}^{n-1}\frac{({\sfrac{1}{2}})_p2^p}{(p+1)!}
C_p^{n-1}
\left\{ {\sfrac{2}{\sqrt{3}}}\
\sinh^{-1}\left({\sfrac{1}{\sqrt{2}}}\right)
+ \sum_{l=1}^p (-1)^l \frac{\Gamma(l)}{({\sfrac{1}{2}})_l\
2^l}\right\}\right] .
\label{Seq5}
\end{align}
Next, generalising Eq.~(\ref{f-eqn}) (Glasser unpublished),  
\begin{align}
&\;_2F_1(-{\sfrac{1}{2}},n;n+{\sfrac{1}{2}};-z) 
\nonumber\\
&\hspace*{2.0cm}
= \left(\frac{1}{2}\right)_n\ 
\left[\frac{1}{n!} + \frac{1}{\sqrt{\pi}\Gamma(n)}
\sum_{k=0}^{n-1}\frac{(-1)^k \Gamma(k+{\sfrac{1}{2}})}{(k+1)!}
\ C^{n-1}_k \left(\frac{z+1}{z}\right)^{k+1}
\right.
\nonumber\\
&\hspace*{3.0cm}\left. \times
\left\{ \sqrt{z}\ \sin^{-1}\left(\sqrt{\frac{z}{z+1}}\right)
- \frac{1}{2} \sum_{l=1}^k \frac{(l-1)!}{({\sfrac{1}{2}})_l}
\ \left(\frac{z}{z+1}\right)^l\right\} \right]
\nonumber\\
&\hspace*{2.0cm}
 = \left(\frac{1}{2}\right)_n \frac{1}{n!}
\left[1 + n(z+1)\ \sin^{-1}\left(\sqrt{\frac{z}{z+1}}\right)
\;_2F_1(1-n,{\sfrac{1}{2}};2;{\sfrac{z+1}{z}})
\right.
\nonumber\\
&\hspace*{5.0cm}\left.
- \frac{n}{2} \sum_{k=0}^{n-1}\sum_{l=1}^k
\frac{({\sfrac{1}{2}})_k\ \Gamma[l]\ (1-n)_l}
{(2)_k\left({\sfrac{1}{2}}\right)_lk!}
\left(\frac{z+1}{z}\right)^{k-l+1}\right] .
\label{Hyper}
\end{align}
The hypergeometric function above is a Jacobi polynomial~\cite{Pr86}
\begin{equation}
\;_2F_1(-N,{\sfrac{1}{2}};2;z)
=\frac{N!}{(2)_N}\ P_N^{(1,-N-\frac{3}{2})}(1-2z) .
\label{Jaceq}
\end{equation}
Therefore, we have our principal closed form result
\begin{align}
&S_k(z)=
\sum_{n=1}^{k+1} n! \left({\frac{z}{4-z}}\right)^n\ E(k,n)
\nonumber\\
&\hspace*{0.3cm}\times
\left[\frac{1}{n} + \sum_{p=0}^{n-1} (-1)^p 
\frac{({\sfrac{1}{2}})_p}{(p+1)!}\ C^{n-1}_p 
\left({\frac{4}{z}}\right)^{p+1}
\left\{ \sqrt{\frac{z}{4-z}}
\sin^{-1}\left({\frac{\sqrt{z}}{2}}\right) 
- \frac{1}{2} \sum_{l=1}^p \frac{\Gamma(l)}{\left({\sfrac{1}{2}}\right)_l}
\left(\frac{z}{4}\right)^l\right\}\right] .
\label{Seq6}
\end{align}
From Eq.~(\ref{Seq6}), {\it for rational} $z$
\begin{equation}
\sum_{m=1}^{\infty}\frac{m^kz^m}{C^{2m}_m}
=R_3(k) + R_4(k)\sqrt{\frac{z}{4-z}}
\sin^{-1}\left(\frac{\sqrt{z}}{2}\right) ,
\label{ratzeq}
\end{equation}
where, again,  $R_j(k)$ is a rational number.

\section{Discussion}
  
As pointed out in the introduction, the radius of convergence for $S_k(z)$ is 
4.  Furthermore, since for large $m$ the asymptotic behavior of the $m-$th term
is $m^{k+1/2}(z/4)^m$, the series converges nowhere on its circle of 
convergence for positive $k$. However Eq.~(\ref{Seq6}) shows that 
$S_k(z)$ is analytic on its two-sheeted Riemann surface joined along a   branch
cut emanating from $z=4$. The corresponding values on the two sheets differ 
only by the sign chosen for $\sqrt{4-z}$ in Eq~(\ref{Seq6}).  For $z=2$ we have
$S_k(2)=R_1(k) + R_2(k)\pi$ where the rational number $R_j(k)$ has denominator 
1 or 2. 
We show values of the ratio,
$R_1(k)/R_2(k)$ and of $S_k(2)$. 
In Table~\ref{table1} (and continued in Table~\ref{t1part2}, we 
list the ratio for $k=1,...,65$ to 65 decimal places.  The last line 
of Table~\ref{t1part2} contains the first 65 decimal digits of the
value of $\pi$.  
For $k=100$, 
96 places of $\pi$ are reproduced. In Table~\ref{table2}, we display the 
first thirty 
complete expressions for $S_k(2)$. As has been demonstrated in the previous 
section, the limit $k\rightarrow\infty$ associated with $S_k(2)$ is indeed 
$\pi$. The rate of convergence is 
derived in the next section.
 
The only previous  non-recursive evaluation of an $S_k$ series appears
to be that of Borwein and Girgensohn~\cite{Bo05} for their
$b_2(k)=S_k(1)$ 
found by a different procedure to ours.  
By comparing their formula with Eq.~\ref{Seq6}) we obtain the intriguing 
identity
\begin{align}
&\sum_{n=1}^{k+1} \frac{(-1)^{n+k+1}n!}{3^n}\ E(k,n)\ C^{2n}_n
\left[\sum_{p=0}^{n-1}
\frac{3^p}{(2p+1)C^{2p}_p} + \frac{2\pi}{\sqrt{27}}\right] 
\nonumber\\
&\hspace*{1.0cm}=
\sum_{n=1}^{k+1} \frac{n!}{3^n} E(k,n]
\left[\frac{2}{n} - \sum_{p=0}^{n-1} (-1)^p 
\frac{\left({\sfrac{1}{2}}\right)_p 4^{p+1}}{(p+1)!} C_p^{n-1}
\left\{ \sum_{l=1}^p\frac{\Gamma(l)}{({\sfrac{1}{2}})_l4^l} 
- \frac{\pi}{\sqrt{27}}\right\}\right] .
\end{align}
Note, that by equating Eqs.~(\ref{Seq4}) and (\ref{Seq6}) for $z=2$, one 
obtains a similar identity.
 
As we wrote at the outset, we were stimulated by Lehmer's paper~\cite{Le85},
and especially his proposal that $S_k(2)$ is intimately related to $\pi$. 
It is true that this way of computing $\pi$ by taking $k\to \infty$ is not as
productive as others, such as Ramanujan's modular equation approach as
given in a nice review~\cite{Ba09}. Nonetheless we believe that the
approach is a 
fascinating contribution to the lure and lore of $\pi$; one which we have 
found both motivating and rewarding. In his honour we have named 
${\lim_{k\to\infty}} R_1(k)/R_2(k)=\pi$,  {\it  Lehmer's limit.}
    
\section{The error}
\label{error}

To derive the error to the $k^{\rm th}$ approximation for $\pi$, $E(k)$, 
we introduce the difference function, $D_k(2)$.
Eq.~(\ref{Seq6}) showed that the function $S_k(z)$ is analytic in the entire
complex plane with a cut along the real axis from 4 to $\infty$.
Furthermore, the function can be continued analytically across the cut
onto other sheets where it is still analytic. It is easy to show that
the sum, $S_k(2) = R_2(k)\ \pi + R_1(k)$, is the value of the function
at the  point $z = 2$ in the first sheet, while the difference, $
D_k(2) = R_2(k)\ \pi - R_1(k)$, is the value of the same function at the
point $z = 2$ on the second sheet.

These two functions define the error,
\begin{equation}
E(k) = \frac{D_k(2)}{R_2(k)} = \frac{2\pi D_k(2)}{S_k(2) + D_k(2)}
\sim \frac{2\pi D_k(2)}{S_k(2)} .
\label{eq1-error}
\end{equation}

We now introduce their generating functions, whose definitions are
\begin{equation}
G(t) = \sum_{k=0}^\infty S_k(2) \frac{t^k}{k!}
\hspace*{0.3cm};\hspace*{0.3cm}
H(t) = \sum_{k=0}^\infty D_k(2) \frac{t^k}{k!} ,
\label{eq2-error}
\end{equation}
where
\begin{equation}
G(t) = \frac{s\left[\sin^{-1}(s) \right]}{\left( 1 -
s^2\right)^{\frac{3}{2}}}  + \frac{1}{\left(1 - s^2\right)} ,
\label{eq3-error}
\end{equation}
and
\begin{equation}
H(t) = \frac{s\left[\cos^{-1}(s) \right]}{\left( 1 -
s^2\right)^{\frac{3}{2}}}  - \frac{1}{\left(1 - s^2\right)} ,
\label{eq4-error}
\end{equation}
with $s = \frac{1}{\sqrt{2}} e^{\frac{t}{2}}$.

It is remarkable that the two functions look similar but have very
different behavior. $G(t)$ has its closest singularity as $s = 1$, with
$t = \ln(2)$, while the closest singularity of $H(t)$ is at $s = -1$,
with $t = \ln(2) + 2\pi i$ and $\ln(2) - 2\pi i$. For large
$k$-values, only these leading singularities need be considered.

Adding Eqs.~(\ref{eq3-error}) and (\ref{eq4-error}) gives
\begin{equation}
R_2(t) = \frac{1}{2} \frac{e^{\frac{t}{2}}}
{\left[2 - e^t \right]^{\frac{3}{2}}} ,
\label{eq5-error}
\end{equation}
and therefore
\begin{equation}
R_2(k) = \frac{k!}{4\pi i} \oint 
\frac{e^{\frac{t}{2}}}{\left(2 - e^t \right)^{\frac{3}{2}}}
\frac{1}{t^{k+1}} dt .
\label{eq6-error}
\end{equation}
After shifting the singularity with $t = \ln(2) + x$, and
introducing the function $g(x)$ which is finite at the origin,
\begin{equation}
g(x) = \left(\frac{x}{e^x - 1}\right)^{\frac{3}{2}}
e^{\frac{x}{2}} = \sum_{j = 0}^\infty c_j x^j ,
\label{eq7-error}
\end{equation}
with a few of the coefficients given by
\begin{displaymath}
\begin{array}{c|rcl}
\hspace*{1.5cm}j\hspace*{1.5cm} &\hspace*{2.5cm} c_j &&\\
\hline
\ 0 & 1 & \\
\ 1 & -1\!&/&\!4\\
\ 2 & -1&/&32\\
\ 3 &  5&/&384\\
\ 4 &  7&/&10240\\
\ 5 & -19&/&40960\\
\ 6 & -869&/&61931520\\
\ 7 &  715&/&49545216\\ 
\ 8 & 2339&/&7927234560\\
\ 9 &-200821&/&475634073600\\ 
10 & -12863&/&1993133260800\\
\hline
\end{array}
\end{displaymath}
we have
\begin{equation}
R_2(k) = \frac{k!}{8\pi i} \oint
\frac{g(x)}{\left(-x\right)^{\frac{3}{2}}
\left\{\ln(2) + x \right\}^{k+1}} dx\ .
\label{eq8-error}
\end{equation}

The contour can be moved to surround the positive $x$-axis
(Hankel contour) and, accommodating the jump across the cut, 
Eq.~(\ref{eq8-error}) becomes
\begin{equation}
R_2(k) = -\frac{k!}{4\pi} \sum_{j=0}^\infty c_j\ 
\left\{\ln(2)\right\}^{\left(j - k -\frac{3}{2}\right)}
\ I_j ,
\label{eq9-error}
\end{equation}
where
\begin{equation}
I_j = \int_0^\infty \frac{x^{\left(j - \frac{3}{2} \right)}}
{\left(1 + x \right)^{k+1}} dx .
\label{eq10-error}
\end{equation}
The integral, $I_j$, can be evaluated with Euler's beta-function
and gamma-function reflection formula, providing the result
\begin{equation}
R_2(k) = \frac{1}{4} \sum_{j=0}^\infty C_j\ 
\left\{\ln(2) \right\}^{j - k -\frac{3}{2}}\ 
\frac{\Gamma\left(k + \frac{3}{2} -j \right)}
{\Gamma\left(\frac{3}{2} - j \right)} ,
\label{eq11-error}
\end{equation}
where $C_j = (-1)^j c_j$.

Similarly, $D_2(k)$ can be obtained, by first observing that 
Eq.~(\ref{eq5-error}) for $R_2(t)$ could have been obtained also by
taking only the first term for $G(t)$ in Eq.~(\ref{eq3-error}), with
$\sin^{-1}(1) = \frac{\pi}{2}$. Thus we have directly from 
Eq.~(\ref{eq4-error}) 
\begin{equation}
D_2(t) = \frac{2\pi e^{\frac{t}{2}}}
{\left[2 - e^t \right]^{\frac{3}{2}}} .
\label{eq12-error}
\end{equation}
Everything goes through in exactly the same way as for $R_2(k)$,
providing the result
\begin{equation}
D_2(k) = -2\pi \Re \left[ 
\sum_{j=0}^\infty C_j 
\left\{\ln(2) + 2\pi i \right\}^{j - k -\frac{3}{2}}
\frac{\Gamma\left(k + \frac{3}{2} -j \right)}
{\Gamma\left(\frac{3}{2} - j \right)} 
\right] .
\label{eq13-error}
\end{equation}

Our final result for the error, $E(k)$, in $\frac{1}{k}$-expansion
form can be obtained readily from Eqs.~(\ref{eq11-error})
and~(\ref{eq13-error}), as
\begin{gather}
E(k) =
\frac{- 8\pi \Re\left[
\sum_{j=0}^\infty F_j 
\left\{\ln(2) + 2\pi i \right\}^{j-k-\frac{3}{2}} 
\prod_{n = 1}^j \frac{1}{(2k + 3 - 2n)}\right]}
{\sum_{j = 0}^\infty F_j \left\{\ln(2) \right\}^{j - k -\frac{3}{2}}
\prod_{n = 1}^j \frac{1}{(2k + 3 -2n)}} ,
\tag{$35a$}
\label{eq14-error}
\end{gather}
where $F_j = -(2j-3)!!\;\; c_j,\ \  (-3)!! = -1,\ {\rm and}\;\; (-1)!! =1
$.  A few of the coefficients are 
\begin{displaymath}
\begin{array}{c|rcl}
\hspace*{1.5cm}j\hspace*{1.5cm} &\hspace*{2.5cm} F_j &&\\
\hline
\ 0 & 1 & \\
\ 1 & 1\!&/&\!4\\
\ 2 &  1&/&32\\
\ 3 & -5&/&128\\
\ 4 & -21&/&2048\\
\ 5 &  399&/&8192\\
\ 6 &  869&/&65536\\
\ 7 & -39325&/&262144\\ 
\ 8 & -334477&/&8388608\\
\ 9 &  28717403&/&33554432\\
10 &   59697183&/&268435456\\
\hline
\end{array}
\end{displaymath}
Note, no Stirling's asymptotic expansion is required in
obtaining Eq.~(\ref{eq14-error}).

The order of magnitude of $E(k)$ can be seen by looking at just the
$j=0$ term of Eq.~(\ref{eq14-error}), which is
\begin{gather}
E_0(k) = - \frac{8\pi}{Q^{k+\frac{3}{2}}} 
\cos\left[\left(k+\frac{3}{2}\right) \phi\right] ,
\notag\\
\intertext{where}
\phi = \tan^{-1}\left( \frac{2\pi}{\ln(2)}\right)
\ =\ 1.4609\ \cdots \ {\rm rad}\ \ = 
83.7047 \cdots \ \deg ,
\tag{$35b$}\\
\intertext{and}
Q = \sqrt{ 1 + \left( \frac{2\pi}{\ln(2)}\right)^2 }
\ =\ 9.1197 \cdots\ \simeq \frac{2\pi}{\ln(2)}  .
\notag
\end{gather}
As $\frac{8\pi}{Q^{\frac{3}{2}}} \simeq 1$, the order of magnitude of
the error is
\begin{gather}
E(k) = {\cal O} \left[\left(\frac{2\pi}{\ln(2)}\right)^{-k}\right]
\hspace*{1.0cm}\left({\rm e.g.\ \ }10^{-96}\ \ {\rm for}\ \ k = 100
\right) .
\tag{$35c$}
\label{eq15-error}
\end{gather}

To assess the overall accuracy of the expansion of $E(k)$ 
given by Eq.~(\ref{eq14-error}), we have compared values 
found from it with just the first seven terms ($j = 0-6$)
with the exact numerical result for $\pi - \frac{R_1(k)}{R_2(k)}$
given in the previous section. That comparison is shown in
Table~\ref{comptab}. 
\begin{table}[h]
\begin{ruledtabular}
\caption{\label{comptab} Values of $E(k)$.}
\begin{tabular}{ccc}
k & Using \ seven\ terms\ in \
Eq.~(\ref{eq14-error}) & exact\ result \\ 
\hline
    5  &      1.394929580622094159D-05 &  1.458979572496761765D-05\\
   10  &      1.208533257871241603D-10 &  1.208426564955768191D-10\\
   15  &     -1.717906040975859856D-15 & -1.717926356677219480D-15\\
   20  &     -5.757361768118465909D-20 & -5.757368219957248525D-20\\
   25  &     -5.054992959778042247D-25 & -5.054994756412840440D-25\\
   30  &      6.170826902775628923D-30 &  6.170826687049165966D-30\\
   35  &      2.283000783226568157D-34 &  2.283000802727741913D-34\\
   40  &      2.208776974546136119D-39 &  2.208776992026544785D-39\\
   45  &     -2.093726806383270815D-44 & -2.093726801667894352D-44\\
   50  &     -9.003884490090622337D-49 & -9.003884494714834077D-49\\
   55  &     -9.605682948187048732D-54 & -9.605682954606388173D-54\\
   60  &      6.756128797397263314D-59 &  6.756128794821975275D-59\\
   65  &      3.530031377831535807D-63 &  3.530031378066361048D-63\\
   70  &      4.137108141343023278D-68 &  4.137108141778060533D-68\\
   75  &     -2.030602678726272460D-73 & -2.030602678498624493D-73\\
   80  &     -1.375355163972183061D-77 & -1.375355163990000725D-77\\
   85  &     -1.764261661388176988D-82 & -1.764261661430246816D-82\\
   90  &      5.380737425072894637D-88 &  5.380737424803436715D-88\\
   95  &      5.323448270450853269D-92 &  5.323448270468033075D-92\\
  100  &      7.454268189222908298D-97 &  7.454268189274244908D-97\\
\hline
\end{tabular}
\end{ruledtabular}
\end{table}
For large values of $k$, the agreement is excellent and that remains 
very good for lower $k$-values. The question of the convergence of the
expansion for $E(k)$ in Eq.~(\ref{eq14-error}), and the continuing 
considerations this question engenders, are discussed in the next
section.
 
\section{Considerations}
 The coefficients $c_j$ behave in an interesting way. The even and odd
coefficients behave differently. They vary as
\begin{displaymath}
c_j \propto [2\pi]^{-j} \left\{
\begin{array}{cc}
\frac{1}{\sqrt{j}} & {\rm for\ even}\ j\\
\sqrt{j}  & {\rm for\ odd}\ j\ \ \\
\end{array}
\right. .
\end{displaymath}
After the first few, each odd-$j$ coefficient is larger than the
preceding even-$j$ one. This occurs because the generating function
for the  series has singularities at plus and minus $2\pi i$ which 
alternately
give terms that  add and subtract. Similar things occur with the Lehmer
errors, $E(k)$, but the periodicity is not so obvious as the phase shift is
not exactly $\frac{\pi}{2}$.

  It is clear now that that the series for $R_2(k)$ is a convergent one,
while that for $D_k(2)$ is divergent, albeit very accurately asymptotic.
It is somewhat daunting to see how commanding the part $2\pi$ plays as
the controlling factor in the error for $\pi$. These
circumstances have motivated our continuing inquiry into the analytic
nature of the error function.

An exact integral representation for the error, $E(k)$, can be obtained.
Returning to Eq.~(\ref{eq6-error}) for $R_2(k)$, and following all of
the procedure thereafter, we find
\begin{equation}
R_2(k) = \frac{(k+1)!}{2\pi}
\int_0^\infty \frac{1}{\sqrt{1 - e^{-x}}
\left\{\ln(2) + x\right\}^{k+2}} dx .
\label{eq16-error}
\end{equation}
Then, using the identity for the gamma function,
\begin{displaymath}
\frac{\Gamma(k+2)}{z^{k+2}} = \int_0^\infty t^{k+1}
\ e^{-zt}\ dt ,
\end{displaymath}
and interchanging orders of integration, gives
\begin{equation}
R_2(k) = \frac{1}{2\sqrt(\pi)} \int_0^\infty x^k\ 
e^{-\ln(2) x}  \frac{\Gamma(x+1)}{\Gamma(x + \frac{1}{2}) } dx .
\label{eq17-error}
\end{equation}
Similarly, and following the exact same procedure, we have an
integral representation for $D_2(k)$, and thus our final formula
for the error,
\begin{equation}
E(k) = - 8\pi \frac
{\int_0^\infty \cos(2\pi x)\ x^k\ e^{-ln(2) x}
\frac{\Gamma(x+1)}{\Gamma(x + \frac{1}{2}) } dx}
{\int_0^\infty  x^k\ e^{-ln(2) x}
\frac{\Gamma(x+1)}{\Gamma(x + \frac{1}{2}) } dx} .
\label{eq18-error}
\end{equation}

To make the connection with this representation for $E(k)$ and that
given in Eq.~(\ref{eq14-error}) we observe that, for large values of
$k$, the major contribution to the integrals comes from large values
of $x$.  Therefore, Stirling's asymptotic expansion can be used to give
\begin{align}
\frac{\Gamma(x+1)}{\Gamma(x + \frac{1}{2}) }
&\sim \sqrt(x) \left[ 1 + \frac{1}{8x} + \frac{1}{128 x^2}
- \frac{5}{1024 x^3} - \frac{21}{32768 x^4} + \cdots \right]
\nonumber\\
&= \sum_{j=0}^\infty A_j\ x^{\frac{1}{2} - j } .
\label{eq19-error}
\end{align}
Inserting this expansion into Eq.~(\ref{eq18-error}), we immediately
recover the series expansion for $E(k)$ given in Eq.~(\ref{eq14-error}) 
where now  $F_j = 2^j A_j$.
To see clearly what this means, we turn to the valuable paper of Tricomi and
Erd$\acute{e}$lyi~\cite{Tr51} (see also Fields~\cite{Fi66}) in which is
given the asymptotic formula for the ratio of gamma functions, namely
\begin{align}
\frac{\Gamma(z+\alpha)}{\Gamma(z+\beta)}
&\sim \sum_{j = 0}^\infty \frac{(-1)^j}{j!} 
\frac{\Gamma(\beta - \alpha +j)}{\Gamma(\beta - \alpha)}
B_j^{(\alpha - \beta +1)}(\alpha)\ x^{(\alpha- \beta -j)} ,\\
&\ \left|arg(z +\alpha) \right| < \pi\ ;\hspace*{1.0cm} 
B_0^{(\alpha - \beta + 1)}(\alpha) = 0 ,
\label{eq20-error}
\end{align}
where the $B_j^{(\alpha - \beta +1)}(\alpha)$ are the generalised
Bernoulli polynomials, the N\"{o}rlund
polynomials~\cite{No24}, defined by
\begin{equation}
\left(\frac{t}{e^t - 1} \right)^\sigma\ e^{xt}
= \sum_{j = 0}^\infty \frac{t^j}{j!} B_j^{(\sigma)}(x) ,
\hspace*{1.0cm} |t| < 2\pi ,
\label{eq21-error}
\end{equation}
and where 
\begin{displaymath}
B_j^{(\sigma)}(\sigma -x) = (-1)^j\ B_j^{(\sigma)}(x) .
\end{displaymath}
The coefficients,$A_j$, and hence the $B_j^{(\sigma)}(x)$, can be
calculated from the formula, Eq.~(11) in ref.~\cite{Tr51},
or they can be computed numerically (as we have done).
Using $\alpha = 1$ and $\beta = \frac{1}{2}$, we find
\begin{displaymath}
A_j = 
-\frac{(2j-3)!!\ (-1)^j\ B_j^{(\frac{3}{2})}(1)}
{2^j j!} = 
-\frac{(2j-3)!!\  B_j^{(\frac{3}{2})}(\frac{1}{2})}
{2^j j!} ,
\end{displaymath}
and comparing the $c_j$ in Eq.~(\ref{eq3-error}) with the
$B_j^{(\sigma)}(x)$ in Eq.~(\ref{eq21-error}) gives
\begin{displaymath}
c_j = \frac{1}{j!} B_j^{(\frac{3}{2})}(\sfrac{1}{2}) .
\end{displaymath}
Therefore, $F_j = 2^j A_j$ is the same as $-(2j - 3)!!\  c_j$.
The connection between the $E(k)$ in the previous section and that
herein is complete.

Tricomi and Erd$\acute{e}$rlyi opened their paper with the salient
comment
``Many problems in mathematical analysis require a knowledge
of the asymptotic behavior of the quotient
$\frac{\Gamma(z_+\alpha)}{\Gamma(z + \beta)}$
for large values of $|z|$.'' There are two, in particular, that bear a
kinship with our work.

Watson~\cite{Wa59} in a lovely cameo entitled {\it A Note on Gamma
Functions}, studied tight bounds on the Wallis formula for $\pi$ by
employing the formula for the hypergeometric function, namely
\begin{displaymath}
\left[\frac{\Gamma(x+1)}{\Gamma(x+\frac{1}{2})} \right]^2 =
x\ F(-\frac{1}{2},-\frac{1}{2};x;1)
= x + \sum_{m=1}^\infty 
\frac{\left( -\frac{1}{2}\right)_m \left( -\frac{1}{2}\right)_m}
{m (x+1)_{m-1} } ,
\end{displaymath}
with the comment, ``the condition $x+\frac{1}{2} > 0$ amply secures the
convergence of the series".

In the classic problem of the one-dimensional random walk~\cite{Woxx},
the expectation value of the absolute distance after $N$ (unit) steps
is given by
\begin{displaymath}
<d_J> = \frac{2}{\sqrt{\pi}} \frac{\Gamma(J+\frac{1}{2})}{\Gamma(J)} ,
\end{displaymath}
where $J = \frac{N}{2}$ for $N$ even, and $\frac{N+1}{2}$ for $N$ odd.
With the asymptotic expansion for the ratio of these gamma functions, we
recover the known result
\begin{displaymath}
<d_N> = \sqrt{\frac{2N}{\pi}} 
\left[1 \mp \frac{1}{4N} + \frac{1}{32 N^2} \pm \frac{15}{128 N^3}
- \frac{21}{2048 N^4} \mp \cdots \right] ,
\end{displaymath}
where the top signs are taken for $N$ even and the bottom ones for $N$
odd.  The expansion for $N$ odd is precisely the one we have for the
$\frac{1}{k}$ expansion for the error, $E(k)$. The alternating signs in
$<d_N>$  is a direct consequence of the $(-1)^j$ phase in
Eq.~(\ref{eq21-error}). We find the apparent, serendipitous,
coincidence ({\it Are There Coincidences In Mathematics?}~\cite{Da81}) 
quite intriguing.

\section{epilogue}

D. H. Lehmer was an eminent mathematician~\cite{Br92}. His many and varied 
works dealt especially with matters of numbers, the queen of mathematics, 
for which he had an abiding and prodigious affection and talent. Many of 
his elegant works bear his name in the literature. Particularly relevant 
here are his Machin-like " On Arccotangent Relations for pi"~\cite{Le38} 
and " A Cotangent Analogue of Continued Fractions"~\cite{Le38a}, in which 
he showed that every positive irrational number has a
unique infinite continued cotangent representation.
 
It would be remiss to close our essay without a salute to $\pi$, the most 
renown of all constants in mathematics. Of the literally plethora of fine 
possible choices, we have chosen to go with,\\
" The value of $\pi$ has engaged the attention of many mathematicians and 
calculators from the time of Archimedes to the present day, and has been 
computed from so many different formulae, that a complete account of its 
calculation would almost amount to a history of mathematics. "\\ 
\hspace*{4.0cm}[J. W. L. Glaisher, Messenger of Math., 25-30 (1872)]\\
and\\ 
" And he made a molten sea, ten cubits from the one brim to the other; it 
was round all about ... and a line of thirty cubits did compass it round 
about."\\
\vskip 0.0001cm
\hspace*{4.0cm}[1 Kings 7:23]\\
\vskip 0.001cm
\hspace*{3.9cm}
[{\it On The Rabbinical Approximation of Pi}~\cite{Bo98}] 

\begin{acknowledgments}
It is with much pleasure and appreciation that we thank our colleague,
and longtime friend, Ken Amos for his assistance with some of the 
numerical work and for his diligent preparation of the manuscript.
\end{acknowledgments}


\begin{table}[h]
\begin{ruledtabular}
\caption{\label{table1} The first 40 values of $R_1(k)/R_2(k)$.}
\begin{tabular}{cl}
1 & 3.0000000000000000000000000000000000000000000000000000000000000000, 
\\
2 & 3.1428571428571428571428571428571428571428571428571428571428571429, 
\\
3 & 3.1428571428571428571428571428571428571428571428571428571428571429, 
\\
4 & 3.1415929203539823008849557522123893805309734513274336283185840708, 
\\
5 & 3.1415780637940682708449916060436485730274202574146614437604924454, 
\\
6 & 3.1415923909792426870850032900640066997667045522521983609499312078, 
\\
7 & 3.1415928223305272485600354452813469206911829862649534780682321666, 
\\
8 & 3.1415926602118958149513766625165465875475594263879799980191580879, 
\\
9 & 3.1415926517076394721680972516575553164503064392560154252064551018, 
\\
10 & 3.1415926534689505819670665641594107024726895639269814592980636856, 
\\
11 & 3.1415926536097034333101766975356113602136332924145393370409221625, 
\\
12 & 3.1415926535916966575323999453816550421791069920332492792677268321, 
\\
13 & 3.1415926535895974772851022724935990443197089788766137673243522672, 
\\
14 & 3.1415926535897658427283911681357586647531232166551257736863740840, 
\\
15 & 3.1415926535897949563890000604989833119363870525182191538675275264, 
\\
16 & 3.1415926535897936078697175611343357401374657772051447922473642141, 
\\
17 & 3.1415926535897932264519119870605439217503823038833322854025505048, 
\\
18 & 3.1415926535897932337393340424095495278361619490164656674686609330, 
\\
19 & 3.1415926535897932384958871798265432629332247495874217158068559674, 
\\
20 & 3.1415926535897932385202170654790753694551295971907030363946530365, 
\\
21 & 3.1415926535897932384636042752252741477211700998636696041572103108, 
\\
22 & 3.1415926535897932384619738974298389685253003898941919111583795322, 
\\
23 & 3.1415926535897932384626159605817069391563876308322608250338128782, 
\\
24 & 3.1415926535897932384626507820001679085203349744076397007956812907, 
\\
25 & 3.1415926535897932384626438887789785254812134010734664381353470044, 
\\
26 & 3.1415926535897932384626433063420855784182255379975982739476954955, 
\\
27 & 3.1415926535897932384626433753704220177671845520815679347756243018, 
\\
28 & 3.1415926535897932384626433840161803084753395271588726611787867400, 
\\
29 & 3.1415926535897932384626433833921567846462444298200187678535709625, 
\\
30 & 3.1415926535897932384626433832733320575101202334085977160126045251, 
\\
31 & 3.1415926535897932384626433832780011313583089291827528827676225744, 
\\
32 & 3.1415926535897932384626433832795412276163249733207758655099465346, 
\\
33 & 3.1415926535897932384626433832795218564970456013498279015329014616, 
\\
34 & 3.1415926535897932384626433832795028764912658671648025552465258989, 
\\
35 & 3.1415926535897932384626433832795026558970891266009144356105970111, 
\\
36 & 3.1415926535897932384626433832795028788309975635209384389235382620, 
\\
37 & 3.1415926535897932384626433832795028868138180463022120372825783062, 
\\
38 & 3.1415926535897932384626433832795028843242936641656711090281944977, 
\\
39 & 3.1415926535897932384626433832795028841687499915165439348592159580, 
\\
40 & 3.1415926535897932384626433832795028841949606223830792761898241293, 
\\
\end{tabular}
\end{ruledtabular}
\end{table}
\begin{table}[h]
\begin{ruledtabular}
\caption{\label{t1part2}
Table~\ref{table1} continued. The $41^{\rm st}$ to 65$^{\rm th}$
values.}
\begin{tabular}{cl}
41 & 3.1415926535897932384626433832795028841974582207305475602005895270, 
\\
42 & 3.1415926535897932384626433832795028841972028735133894452619084772, 
\\
43 & 3.1415926535897932384626433832795028841971667284657301213102341152, 
\\
44 & 3.1415926535897932384626433832795028841971689329076588332348800482, 
\\
45 & 3.1415926535897932384626433832795028841971694203123738376538881191, 
\\
46 & 3.1415926535897932384626433832795028841971694054855024473294830381, 
\\
47 & 3.1415926535897932384626433832795028841971693992698167077733580624, 
\\
48 & 3.1415926535897932384626433832795028841971693992991125354048682206, 
\\
49 & 3.1415926535897932384626433832795028841971693993745498910645297440, 
\\
50 & 3.1415926535897932384626433832795028841971693993760062094244160757, 
\\
51 & 3.1415926535897932384626433832795028841971693993751340966582560562, 
\\
52 & 3.1415926535897932384626433832795028841971693993750956731257005236, 
\\
53 & 3.1415926535897932384626433832795028841971693993751052376002021699, 
\\
54 & 3.1415926535897932384626433832795028841971693993751059289957208672, 
\\
55 & 3.1415926535897932384626433832795028841971693993751058305806275469, 
\\
56 & 3.1415926535897932384626433832795028841971693993751058199065942028, 
\\
57 & 3.1415926535897932384626433832795028841971693993751058208338153750, 
\\
58 & 3.1415926535897932384626433832795028841971693993751058209844036282, 
\\
59 & 3.1415926535897932384626433832795028841971693993751058209768684753, 
\\
60 & 3.1415926535897932384626433832795028841971693993751058209748770310, 
\\
61 & 3.1415926535897932384626433832795028841971693993751058209749198387, 
\\
62 & 3.1415926535897932384626433832795028841971693993751058209749448105, 
\\
63 & 3.1415926535897932384626433832795028841971693993751058209749448952, 
\\
64 & 3.1415926535897932384626433832795028841971693993751058209749445970, 
\\
65 & 3.1415926535897932384626433832795028841971693993751058209749445888
\\
\hline
$\pi$ & 3.1415926535897932384626433832795028841971693993751058209749445923
\\
\end{tabular}
\end{ruledtabular}
\end{table}
 
\begin{table}[h]
\begin{ruledtabular}
\caption{\label{table2}$S_k(2)$ for $k=1,\dots,30$}
\begin{tabular}{cl}
1 & 3 + $\pi$\\
2 & 11 + (7 $\pi$)/2\\
3 & 55 + (35 $\pi$)/2\\
4 & 355 + 113 $\pi$\\
5 & 2807 + (1787 $\pi$)/2\\
6 & 26259 + (16717 $\pi$)/2\\
7 & 283623 + 90280 $\pi$\\
8 & 3473315 + (2211181 $\pi$)/2\\
9 & 47552791 + (30273047 $\pi$)/2\\
10 & 719718067 + 229093376 $\pi$\\
11 & 11932268231 + (7596317885 $\pi$)/2\\
12 & 215053088835 + (136907048461 $\pi$)/2\\
13 & 4186305575415 + 1332542451241 $\pi$\\
14 &  87534887434835 + (55726440112987 $\pi$)/2\\
15 & 17/2 (230197719678574 + 73274209950655 $\pi$)\\
16 &  46561960552921315 + 14821132364094533 $\pi$\\
17 & 1175204650272267479 + (748158516941653967 $\pi$)/2\\
18 & 31357650670190565363 + (19962900431638852297 $\pi$)/2\\
19 & 881958890078887314567 + 280736233919792968780 $\pi$\\
20 & 26078499305918584929155 + (16602088291822017588121 $\pi$)/2\\
21 & 808742391638178302137783 + (514861397268710391722627 $\pi$)/2\\
22 & 26247592141035336332994451 + 8354868067011516415979216 $\pi$\\
23 & 889735042691243752903048295 + (566422920345559866343383785 $\pi$)/2\\
24 & 
31443867356631172742458654755 + (20017787678934958873836057001 $\pi$)/2\\
25 & 1156619309474553778799639807127 + 
368163360756819772832459706481 $\pi$\\
26 & 44213527064791762795003086702899 + 
  (28147205535555628918615623800767 $\pi$)/2\\
27 &  29/2 (120960271409361525575952065166694 + 
    38502850225074296050584562759015 $\pi$)\\
28 &  72107782245849606090651464405624515 + 
  22952619959641949809923983081211353 $\pi$\\ 
29 &  3068555154632012211023759300893608311 + 
  (1953502884039199982724152729710551347 $\pi$)/2\\
30 &  135010171084427194623890031993168567507 + 
  (85950144383076253408132013000868398677 $\pi$)/2\\
\end{tabular}
\end{ruledtabular}   
\end{table} 

\end{document}